\documentclass[12pt]{article}

\usepackage{soul}

\usepackage{tikz}
\usepackage{epsfig,latexsym,amsfonts,amsmath,amsthm,amssymb,amsbsy,multirow,slashed,wasysym,textcomp,subfigure,wrapfig,datetime,comment,mathtools,cancel,cite,twistor,mathrsfs}
\usepackage[hidelinks]{hyperref}
\usetikzlibrary{calc}
\usepackage[font={footnotesize},bf]{caption}

\usepackage[normalem]{ulem} 
\numberwithin{equation}{section}
\def\ee{\end{equation}}
\def\be{\begin{equation}}
\def\bea{\begin{eqnarray}}
\def\eea{\end{eqnarray}}
\newcommand{\beq}{\begin{eqnarray}}
\newcommand{\eqq}{\end{eqnarray}}
 \newcommand{\badat}{\begin{alignedat}}
 \newcommand{\eadat}{\end{alignedat}}

\newcommand{\eal}[1]{\be \begin{aligned} #1 \end{aligned}\end{equation}} 

\newcommand{\eqn}[1]{\be #1 \end{equation}} 
\newcommand{\eqa}[1]{\bea  #1\end{eqnarray}}

\newcommand{\avg}[1]{\left< #1 \right>} 
\long\def\new#1\endnew{{\bf #1}}		
\long\def\del#1\enddel{}

\def\del{\partial}

\def\s{\sigma }

\usepackage{color}

\newcommand{\pink}[1]{\textcolor{\pink}{#1}}

\definecolor{dblue}{rgb}{0.2,0.50,0.80}

\usepackage{xspace}
\def\C{\mathcal{C}}

\def\N{\mathcal{N}}

\def\M{\mathcal{M}}

\newcommand{\ket}[1]{\left|#1\right\rangle}
\newcommand{\bra}[1]{\left\langle#1\right|}

\def\bz{{\bar z}}

\def\s{ {\sigma} }

\def\scri{{\mathscr{I}}}
\def\t2{T$^{1,1}$}

\newcommand{\btau}{\bar{\tau}}
\catcode`,\active

\catcode`\,12

\newcommand{\sumint}{%
  \mathop{%
    \ooalign{%
      $\displaystyle\sum$\cr
      \hidewidth$\displaystyle\int$\hidewidth\cr
    }%
  }%
}

\newcommand{\lket}[2]{\ensuremath{\left|#2\right\rangle_{#1}}}
\newcommand{\lbra}[2]{\ensuremath{{}_{#1}\left\langle#2\right|}}
\newcommand{\lbraket}[4]{\ensuremath{{}_{#1}\left\langle#3\right|\left.#4\right\rangle_{#2}}}
\newcommand{\inte}{\mathrm{in}}
\newcommand{\out}{\mathrm{out}}

\begin{document}
\begin{titlepage}
\unitlength = 1mm~\\
\vskip 3cm
\begin{center}

{\LARGE{Observing Massive Scattering from Null Infinity}}

\vspace{0.8cm}
Walker Melton$^{ab}$\footnote{wmelton@fas.harvard.edu}, Kyrill Michaelsen$^c$\footnote{kyrill.michaelsen@uni-hamburg.de}, and Romain Ruzziconi$^{ade}$\footnote{romainruzziconi@fas.harvard.edu}\\
\vspace{1cm}

{\it  $^a$Center for the Fundamental Laws of Nature, Harvard University, Cambridge, MA 02138, USA
\newline
$^b$Society of Fellows, Harvard University, Cambridge, MA 02138, USA 
\newline  
$^c$Department of Mathematics, University of Hamburg, Hamburg, Germany 
\newline 
$^d$Black Hole Initiative, Harvard University, Cambridge, MA 02138, USA 
\newline 
$^e$Centre de Physique Théorique, École polytechnique,
91120 Palaiseau, France
} \\

\vspace{0.8cm}

\begin{abstract}
Because massive particles asymptote to timelike rather than null infinity, current flat space holographic proposals such as celestial or Carrollian holography struggle to describe scattering processes with massive external states. We take a step toward addressing this limitation by studying how information about massive scattering amplitudes is carried to the late-time limit of null infinity by soft graviton radiation. We show that continuity between the boundaries of timelike and null infinity implies that the late-time limit of the Bondi mass aspect naturally acts as a detector operator for massive outgoing radiation. We further relate in-in correlation functions of the Bondi mass aspect at $\mathscr{I}^+_+$ to weighted sums of scattering cross sections, implying that an observer at $\mathscr{I}$ can extract information about massive scattering processes at late times. Finally, we interpret the Bondi mass aspect as a Carrollian stress-tensor component, and study Ward identities to constrain its two-point functions.

 \end{abstract}

\end{center}

\end{titlepage}

\tableofcontents
\section{Introduction}

The AdS/CFT correspondence has provided an invaluable tool for studying string theory and quantum gravity in asymptotically anti-de Sitter spacetimes by recasting gravitational physics as a lower-dimensional CFT living on the boundary of spacetime. Extending holographic dualities to more general contexts, such as asymptotically-flat or de Sitter spacetimes, therefore promises to give us tools to study quantum gravity in more general and physically realistic regimes. 

To date, the holographic dictionary for asymptotically flat spacetimes has focused on recasting bulk scattering amplitudes as correlation functions of a putative dual theory living at the conformal boundary of spacetime. In celestial holography, the dual `celestial CFT' would live on the codimension-2 celestial sphere \cite{Pasterski:2016qvg,Pasterski:2021raf}. In Carrollian holography the dual is a Carrollian CFT that would live on the entirety of null infinity \cite{Ruzziconi:2026bix}. 

These dictionaries are natural for massless scattering, where external particles naturally begin and end on null infinity $\mathscr{I}$, the conformal boundary of asymptotically flat spacetimes. In celestial holography, Mellin transforms map momentum-space scattering amplitudes to celestial correlation functions that could be computed by a putative dual CFT \cite{Pasterski:2017kqt, Pasterski:2017ylz}. Universal collinear divergences allow the study of singular terms of massless-massless operator product expansions \cite{Pate:2019lpp}. Operators with special conformal dimensions encode universal factorization properties of amplitudes when massless legs become soft \cite{Pate:2019mfs, Puhm:2019zbl, Donnay:2018neh}; combining these implies the presence of intricate infinite dimensional symmetry algebras that should significantly constrain the form of the dual theory \cite{Guevara:2021abz, Strominger:2021mtt}. In special cases, explicit dualities relating forms of self-dual gravity and gauge theories on nontrivial backgrounds have been constructed \cite{Costello:2022jpg, Costello:2023hmi}.

The Carrollian picture for massless scattering is similarly well developed. Integral transforms relating momentum-space amplitudes to correlation functions of Carrollian primary operators are known \cite{Donnay:2022aba, Bagchi:2022emh, Donnay:2022wvx,Mason:2023mti}. The imprints of the leading and subleading gravitational soft algebra on the hypothetical Carrollian dual have been computed \cite{Ruzziconi:2024kzo,Kraus:2025wgi,Isen:2026xoc}. Carrollian correlators have been obtained from the flat-space/Carrollian limit of AdS/CFT boundary correlation functions \cite{Bagchi:2023fbj, Alday:2024yyj, deGioia:2024yne,Lipstein:2025jfj, Kraus:2024gso, Kulkarni:2025qcx ,  Marotta:2025qjh, Adamo:2025bfr}. Furthermore, tree-level Yang-Mills amplitudes have been shown to be reproduced from an effective Carrollian action living at $\mathscr{I}$ \cite{Opreij:2026bdx}, raising interesting questions on the quantization and locality of such dual theories \cite{deBoer:2023fnj, Cotler:2024xhb, Cotler:2025dau, Cotler:2025npu}.

However, because massive trajectories always begin and end at timelike infinity, both approaches fail to naturally include massive states. In celestial holography, massive conformal primary wavefunctions are known. However, the absence of collinear factorizations obscures the structure of massive celestial OPEs, and these amplitudes exhibit features that cannot be constructed by any local conformal field theory \cite{Himwich:2023njb}. In Carrollian holography, local Carrollian primary field representations are necessarily massless (see, however, \cite{Kulp:2024scx,Nguyen:2025sqk} for a discussion on massive Carrollian multiplets and \cite{Ruzziconi:2026isv} for massive representations in terms of extended objects). Attempts to encode massive fields at null infinity have recently been discussed in \cite{Ammon:2025avo} using a holographic renormalization procedure for a massive scalar field, and in \cite{Liu:2026toc} using a complex momentum shift to build a massive Carrollian wavefunction. Furthermore, massive fields near timelike infinity can be reinterpreted in terms of Carrollian physics \cite{Have:2024dff}, but a conventional codimension-one holographic dictionary for massive scattering amplitudes does not naturally emerge from this framework.

Despite this, any consistent dual to flat space quantum gravity must clearly include massive external states; not only are they present in the real world, they appear to be necessary for string theory's internal consistency. Understanding how the celestial or Carrollian theory describes massive scattering amplitudes is therefore critical for constructing a full holographic duality for asymptotically flat spacetimes. 

In this work, we take a first step to understanding how the dual theory incorporates massive scattering by asking a simpler question: how does information about massive scattering amplitudes reach null infinity? While in a purely massive theory the $S$-matrix will not be accessible from measurements at $\mathscr{I}$, gravitational effects should carry information about massive scattering processes to $\mathscr{I}$. 

So far, in the Carrollian holography picture, only radiative data of the gravitational field have been identified with massless bulk scattering states. However, it is well known that null infinity contains another piece of data: the Carrollian stress tensor, which encodes information about the mass and angular momentum aspects of the bulk gravitational solution \cite{Donnay:2022aba, Fiorucci:2025twa , Hartong:2025jpp}. In contrast with AdS/CFT, this object is not associated with bulk graviton scattering states, which are instead encoded in the asymptotic shear. This naturally suggests studying whether the Carrollian stress tensor at null infinity encodes information about massive scattering. In this paper, we show that this is indeed the case, using matching conditions between massive fields at $i^\pm$ and the components of the Carrollian stress tensor at $\mathscr{I}_\pm^\pm$.

To understand how information about the massive $S$-matrix can be observed at $\mathscr{I}$, we study more general asymptotic observables. Beyond familiar scattering amplitudes, quantum field theories and quantum gravity possess a rich set of asymptotic observables \cite{Caron-Huot:2023vxl, Herrmann:2024yai, Caron-Huot:2023ikn}. In particular, we focus on in-in observables, which calculate expectation values of operators in the presence of a specified ingoing state. In-in correlation functions of so-called detector operators, such as the energy/ANEC operator for massless particles, have been used to relate QFT predictions to collider experiments \cite{Herrmann:2024yai}. They also provide a useful class of IR finite observables in gauge theory and gravity \cite{Herrmann:2024yai, Gonzalez:2025ene, Moult:2025njc}.

In this paper we identify a detector operator for massive external particles that naturally lives at null infinity. While outgoing massive particles always reach timelike infinity $i^+$ rather than null infinity $\mathscr{I}$, soft graviton radiation  transfers information about the outgoing massive state to null infinity, allowing observers sitting on null infinity to infer information about the massive $S$-matrix by measuring long-wavelength gravitational radiation at late time. In particular, we show that in-in correlation functions of the Bondi mass aspect, interpreted as a component of the Carrollian stress tensor at $\mathscr{I}$ \cite{Donnay:2022aba, Fiorucci:2025twa , Hartong:2025jpp}, are related to averages of massive $S$-matrix elements. Our proposal differs from that in \cite{Have:2024dff} as it can naturally be measured at $\mathscr{I}^+_+$ and measures a weighted average of massive flux pointing in every direction of timelike infinity.

This paper is organized as follows. In Section \ref{sec:background}, we briefly review metrics for asymptotically flat spacetimes near null and timelike infinity and how continuity across $\mathscr{I}^+_+$ relates subleading terms in the metric.  In Section \ref{sec:sourcecarroll}, we show that the late-time limit of the stress tensor of a massive field sources the late time limit of the Bondi mass aspect, which is proportional to a component of the Carrollian stress tensor at $\mathscr{I}^+_+$ and effectively operates as a detector operator for massive particles that can be measured on $\mathscr{I}$. In Section \ref{sec:Carrstress}, we discuss general properties of the Carrollian stress tensor correlation functions using Ward identities in the Carrollian CFT at $\mathscr{I}$. In Section \ref{sec:inin}, we show that in-in correlation functions of this late time limit of the Carrollian stress tensor can be  expressed as averages of massive scattering amplitudes. Finally, in Section \ref{sec:apps} we compute low-point in-in correlators of the Carrollian stress tensor at $\mathscr{I}^+_+$ in the presence of one- and two-particle ingoing massive wavepackets. Appendix \ref{sec: Appendix Boundary correlators} contains details on the derivation of the Carrollian stress tensor two-point functions and Appendix \ref{app:twopoint} contains details of the calculation of the one- and two-point in-in correlators of our detector operator in a two-particle ingoing state.  

\section{Matching conditions}\label{sec:background}

In this section, we briefly review the Beig-Schmidt expansions near $i^\pm$, the Bondi expansions near $\mathscr{I}^\pm$, and the matching conditions between them. We mostly follow \cite{Compere:2023qoa} (see also \cite{Compere:2011ve,Troessaert:2017jcm,Henneaux:2018cst,Prabhu:2019fsp,Capone:2022gme,Compere:2026jmk, Boschetti:2026gfd}).

\subsection{Metric expansions in asymptotically flat spacetimes}

\subsubsection{Future timelike infinity $i^+$}

Near $i^+$, Cartesian coordinates $X^\mu$ in four-dimensional Minkowski space can be written in terms of the Beig-Schmidt coordinates $(\tau, \rho, z,\bz)$ as
\begin{equation} \label{BS coord}
    X^\mu = \tau \hat{x}^\mu, \qquad \hat{x}^\mu = (\cosh\rho,\sinh\rho \, \vec{n}), \qquad \vec{n} = \frac{1}{1 + z\bar{z}} \left( z + \bar{z}, -i(z - \bar{z}), 1 - z\bar{z} \right),
    \end{equation}
where $\hat{x}^\mu \eta_{\mu\nu}\hat{x}^\nu = -1$ and $\vec{n} \in S^2$. The vector $\hat{x}^\mu$ is supported on the unit Euclidean AdS$_3$ hyperboloid parametrized by $\varphi^a = (\rho, z, \bar z)$. In these coordinates, the Minkowski metric takes the form 
    \begin{equation}
        ds^2 = \eta_{\mu\nu} dX^\mu d X^\nu  = -d\tau^2 + \tau^2 h_{ab}d\varphi^ad\varphi^b,\qquad h_{ab}d\varphi^ad\varphi^b = d\rho^2 + \sinh^2\rho d\Omega_2^2,
    \end{equation}
where $h_{ab}$ is the metric on the three-dimensional hyperboloid and $d\Omega_2^2 = \frac{4 \, dz \, d\bar{z}}{(1 + z\bar{z})^2}$ the metric on $S^2$.

The point $i^+$ can be reached by sending $\tau \to \infty$. In this limit, a general asymptotically flat metric can be expanded as \cite{Compere:2023qoa} 
\begin{equation}
\begin{split}
    ds^2 = &\left(-1-\frac{2\sigma}{\tau} - \frac{\sigma^2}{\tau^2} + o(\tau^{-2})\right)d\tau^2 + o(\tau^{-1}) d\tau d\varphi^a \\
    &+ \tau^2\left(h_{ab}+\tau^{-1}(k_{ab}-2\s h_{ab}) + \tau^{-2}\log\tau i_{ab} + \tau^{-2}j_{ab} + o(\tau^{-2})\right)d\varphi^a d\varphi^b,
\end{split}
\end{equation} where $\sigma$, $k_{ab}$, $i_{ab}$, and $j_{ab}$ are functions of the coordinates on the hyperboloid. In vacuum, the mass aspect $\sigma$ obeys
\begin{equation}
    R_{\tau\tau} - \frac{1}{2}g_{\tau\tau}R = \frac{1}{\tau^3}2(\hat{\nabla}^2-3)\s +{O}\left(\frac{\log\tau}{\tau^{4}}\right) = 0,
    \label{vacuum eq sigma}
\end{equation}
where $\hat{\nabla}^2$ is the Laplacian of the metric $h_{ab}$. 
\subsubsection{Past timelike infinity $i^-$}

The description of past timelike infinity can be obtained from the metric of future timelike infinity by changing $\tau\rightarrow-\btau$ where $\btau = -\sqrt{t^2-r^2}$. This gives 
\begin{equation}
\begin{split}
    ds^2 = &\left(-1+\frac{2\sigma}{\btau} - \frac{\sigma^2}{\btau^2} + o(\btau^{-2})\right)d\btau^2 + o(\btau^{-1}) d\btau d\varphi^a \\
    &+ \btau^2\left(h_{ab}-\btau^{-1}(k_{ab}-2\s h_{ab}) + \btau^{-2}\log(-\btau) i_{ab} + \btau^{-2}j_{ab} + o(\btau^{-2})\right)d\varphi^a d\varphi^b .
\end{split}
\end{equation}
In vacuum, $\s$ will obey 
\begin{equation}
   R_{\btau\btau}-\frac{1}{2}g_{\btau\btau}R =-\frac{1}{\bar \tau^3} 2 (\hat{\nabla}^2-3)\s +{O}\left(\frac{\log|\bar \tau|}{\bar\tau^{4}}\right) = 0
\end{equation}
near $i^-$. 

\subsubsection{Future null infinity $\mathscr{I}^+$}

The asymptotically flat region near future null infinity is parametrized by retarded Bondi coordinates: $u$, $r$, $x^A$, with $x^A = (z, \bar z)$. They are related to Beig-Schmidt coordinates by
\begin{equation}
    u=\tau e^{-\rho}, \qquad r=\tau \,\sinh\,\rho .
\end{equation}
In these coordinates, null infinity is reached by taking $r\to + \infty$, and an asymptotically flat metric becomes  
\begin{equation}
    \begin{split}
    &ds^2 = 
\left( -1 + \frac{2m_B}{r} + O(r^{-2}) \right) du^2 
+ 2 \left( -1 + \frac{1}{16r^2}C_{AB}C^{AB}  + O(r^{-3})\right) du dr\\
&+ 2r \left( \frac{1}{2r} \nabla^B C_{AB} + \frac{2}{3r^2} \left( N_A + u\partial_A m_B - \frac{3}{32}\partial_A (C_{BC} C^{BC}) \right) + O(r^{-3}) \right) du dx^A \\
&+ r^2\left( \gamma_{AB} + r^{-1} C_{AB} + O(r^{-2}) \right) dx^A dx^B,
\end{split}
\end{equation} where $\gamma_{AB}dx^A dx^B = d\Omega^2_2$, $C_{AB}$ is the asymptotic shear, $m_B$ the Bondi mass aspect, and $N_A$ the angular momentum aspect. These are functions of the coordinates $(u,z, \bar z)$ at $\mathscr{I}^+$.

\subsubsection{Past null infinity $\mathscr{I}^-$}

The advanced Bondi expansion near $\mathscr{I}^-$ can be obtained from the retarded Bondi expansion near $\mathscr{I}^+$ by sending $u \to - v$. We have explicitly 
\begin{equation}
    \begin{split}
    &ds^2 = 
\left( -1 + \frac{2m_B}{r} + O(r^{-2}) \right) dv^2 
- 2 \left( -1 + \frac{1}{16r^2}C_{AB}C^{AB} + O(r^{-3}) \right) dv dr\\
&- 2r \left( \frac{1}{2r} \nabla^B C_{AB} + \frac{2}{3r^2} \left( N_A - v\partial_A m_B - \frac{3}{32}\partial_A (C_{BC} C^{BC}) \right) + O(r^{-3}) \right) dv dx^A \\
&+ r^2\left( \gamma_{AB} + r^{-1} C_{AB} + O(r^{-2}) \right) dx^A dx^B,
\end{split}
\end{equation} and the terms appearing in the expansion are now functions of the coordinates $(v, z, \bar z)$ at $\mathscr{I}^-$.

\subsection{Matching $i^\pm$ with $\mathscr{I}^\pm$}
\label{Matching section}

The matching condition relating the mass aspect $\sigma$ at $i^+$ to the mass aspect $m_B$ at $\mathscr{I}^+$ is given by \cite{Compere:2023qoa}:
\begin{equation}\label{matching}
    \sigma=-2m^{(0)}_Be^{-3\rho}+o(e^{-4\rho}),
\end{equation} where $m^{(0)}_B$ is the leading order of the expanded Bondi mass aspect around $\mathscr{I}^+_+$:
\begin{equation}
    m_B=m_B^{(0)}+m_B^{(1)}u^{-1}+o(u^{-1}).
\end{equation} Hence, the leading order in the $\rho$-expansion of $\sigma$ on the hyperboloid coincides with the value of the Bondi mass aspect at $\mathscr{I}^+_+$. A similar matching can be achieved for the other quantities appearing in the metric expansions, and we refer to \cite{Compere:2023qoa} for details. To perform an analogous matching of $i^-$ with $\mathscr{I}^-$, we just identify the expansion of the mass aspect around $\mathscr{I}^-_-$:
\begin{equation}
    m_B(v)=m_B^{(0)}-m^{(1)}_B v^{-1}+o(v^{-1}).
\end{equation}
From this one can derive the same matching of $m^{(0)}_B$ with $\sigma$ as in Equation \eqref{matching}.

\section{Sourcing the Bondi mass aspect with massive fields}\label{sec:sourcecarroll}

Our objective is to understand which information about massive scattering can be holographically encoded at null infinity. We focus on massive Klein-Gordon scalar fields. In this section, we explain how such fields can be probed in terms of operators of the linearized gravitational field around Minkowski spacetime.

\subsection{Massive encoding at $i^\pm$}

Consider a free massive scalar field $\phi(x)$ with mass $m$. When massive flux passes through $i^\pm$, $\sigma$ will be sourced by the leading $T_{\tau \tau}$ component of the stress tensor. For brevity, we only write the expressions at $i^+$, but completely analogous expansions hold at $i^-$. Near $i^+$, the Fourier mode expansion of the field reads as \cite{Campiglia:2015kxa, Agrawal:2023zea , Have:2024dff, Ruzziconi:2026isv}
\begin{equation}\label{KGfield}
\phi^{(\mathrm{free})}(\tau, p) 
= \frac{\sqrt{m}}{2 (2\pi \tau)^{3/2}}
\left( b(p)\, e^{-i \tau m} + b(p)^{\dagger} e^{i \tau m} \right)
+ O(\tau^{-5/2}),
\end{equation}
where $b(p)^\dagger$, $b(p)$ are the creation/annihilation operators of the field and their momentum $p$ is parametrized as 
\begin{equation}\label{eq: momentum}
    p^\mu(\rho,\vec{n}) = m\hat x^\mu=m(\cosh\rho,\sinh\rho \, \vec{n}), 
\end{equation} and the Beig-Schmidt coordinates $(\tau, \varphi^a)$, $\varphi^a = (\rho, z, \bar z)$, were introduced in \eqref{BS coord}.

The stress tensor of the massive Klein-Gordon field is of the form:
\begin{equation}
    T_{\mu\nu}=\partial_\mu\phi\partial_\nu\phi-\frac{1}{2}g_{\mu\nu}(\partial^{\alpha}\phi\partial_\alpha\phi+m^2\phi^2).
\end{equation}
Near $i^+$, the large $\tau$ expansion of the $T_{\tau\tau}$ component reads as 
\begin{equation}
    T_{\tau\tau}(\tau,\hat{x})= \frac{m^3}{4(2\pi\tau)^3}\left[b(m\hat{x})b(m\hat{x})^\dagger+b(m\hat{x})^\dagger b(m\hat{x})\right]+O(\tau^{-4}) . 
\end{equation}
In the quantum theory, we will need to normal order the expression, giving 
\begin{equation} \label{modes massive}
     T_{\tau\tau}(\tau,\hat{x})= \frac{1}{\tau^3}T^{(3)}_{\tau\tau}(\hat{x}) +O(\tau^{-4}) =  \frac{m^3}{2(2\pi\tau)^3}\left[b(m\hat{x})^\dagger b(m\hat{x}) + a\right]+O(\tau^{-4}) .
\end{equation}
Requiring that $T_{\tau\tau}$ annihilate the vacuum requires the constant $a$ to vanish, $a = 0$. 
Inserting the large $\tau$ expansion of the stress tensor into Einstein's equation implies that 
\begin{equation}    
(\hat \nabla^2-3)\sigma=4\pi G_NT^{(3)}_{\tau\tau},
\end{equation} which is the sourced version of \eqref{vacuum eq sigma} at order $O(\tau^{-3})$. This can be solved as
\begin{equation}\label{eq: sigma}
    \sigma(\hat{x}) = 4\pi G_N\int d^3\hat{x}'\sqrt{h} G_{BB}(\hat{x};\hat{x}')T^{(3)}_{\tau\tau}(\hat{x}'),
\end{equation}
where $G_{BB}$ is the hyperboloid bulk-to-bulk propagator \cite{Campiglia:2015lxa} obeying 
\begin{equation}\label{eq: Green's function}
    (\hat \nabla^2-3)G_{BB}(\hat{x};\hat{x}') = \delta^{(3)}(\hat{x},\hat{x}') ,
\end{equation}
where the $\delta$-distribution is defined so that $\int d^3\hat{x}\sqrt{h}f(\hat{x})\delta^{(3)}(\hat{x}',\hat{x}) = f(\hat{x}')$ and the measure takes the form: $d^3\hat{x}\sqrt{h}=\sinh^2{\rho}\,d\rho\, d\Omega_2$. Hence, Equation \eqref{eq: sigma} shows that the operator $\sigma$ encodes the insertion of a stress tensor component $T_{\tau \tau}$ at $i^\pm$. We can now relate this operator to the Bondi mass aspect $m_B$ at $\mathscr{I}^\pm_\pm$ via the matching conditions discussed in Section \ref{Matching section}. Similar gravitational operators could be built from the insertion of the other stress tensor components at $i^\pm$, which we leave for future analysis. 

\subsection{Massive encoding at $\mathscr{I}^\pm_\pm$}
\label{sec:Massive encoding at I}

Following Section \ref{Matching section}, continuity across $\scri^+_+$ implies that the late limit of the Bondi mass aspect is 
\begin{equation}
    \begin{split}
        m_B^{(0)}(\vec{n}) &= \lim_{u\to \infty} m_B(u,\vec{n}) \\
        &= -\frac{1}{2}\lim_{\rho \to \infty}e^{3\rho}\sigma(\rho,\vec{n}) \\
        &= -2\pi G_N\int d^3\hat{x}\sqrt{h}G_{Bb}(\hat{x};\vec{n})T^{(3)}_{\tau\tau}(\hat{x}),
    \end{split}
\end{equation}
where 
\begin{equation}
    G_{Bb}(\hat{x};\vec{n}) = \lim_{\rho \to \infty} e^{3\rho}G_{BB}(\rho,\vec{n};\hat{x})
\end{equation}
is a bulk-to-boundary propagator.\footnote{Our convention for the normalization for the bulk-to-boundary propagator may differ by an overall constant from others in the literature.} In terms of the modes of the massive field in \eqref{modes massive}, this becomes
\begin{equation}
    m_B^{(0)}(\vec{n}) = -\frac{m^3G_N}{8\pi^2}\int d^3\hat{x}\sqrt{h} G_{Bb}(\hat{x};\vec{n}) b(m\hat{x})^\dagger b(m\hat{x}) . 
\end{equation}
For convenience, we define the rescaled operator 
\begin{equation} \label{matching M with T}
\begin{split}
    \mathcal{M}(\vec{n}) = \frac{1}{4\pi G_N}m_B^{(0)}(\vec{n}) &= -\frac{1}{2}\int d^3\hat{x}\sqrt{h} G_{Bb}(\hat{x};\vec{n})T^{(3)}_{\tau\tau}(\hat{x}) \\ 
    &= -\frac{m^3}{32\pi^3}\int d^3\hat{x}\sqrt{h}G_{Bb}(\hat{x};\vec{n})b(m\hat{x})^\dagger b(m\hat{x}).
\end{split}
\end{equation}
We choose our normalization so that the leading term in correlation functions of $\mathcal{M}$, corresponding to diagrams with no unobserved or exchanged gravitons, is at order $\mathcal{O}(G_N^0)$. Since the creation and annihilation operators are normal ordered, any correlator of $\mathcal{M}$ vanishes in the vacuum, as will be confirmed in the next section by a boundary analysis.

\section{Ward identities for the Carrollian stress tensor}
\label{sec:Carrstress}

The rescaled Bondi mass aspect $\mathcal{M}$, introduced in the previous section, can naturally be interpreted as the $uu$-component (in retarded Bondi coordinates) of the stress tensor of a conformal Carrollian field theory living at null infinity \cite{Donnay:2022wvx, Fiorucci:2025twa ,Hartong:2025jpp}. This motivates the following analysis of the conformal Carrollian stress tensor and its correlators from a boundary perspective. We begin this section by providing a quick review of the symmetries of the Carrollian geometry underlying the conformal Carrollian field theory at $\mathscr{I}$ and analyze the symmetries of its stress tensor multiplets. We determine the two-point functions of the stress tensor components (i.e. of the Carrollian momenta), and show the consistency of this analysis with the previous section. 

\subsection{Conformal Carrollian symmetries}

As a null hypersurface, null infinity naturally admits a Carrollian structure, which consists of a degenerate metric $q_{ab}$ and a nowhere vanishing vector field $n^c$ that lies in the kernel of $q_{ab}$ with signature $(0,+,+)$, i.e. satisfying:
\begin{equation}
    q_{ac}n^c=0 .
\end{equation} At $\mathscr{I}$, this geometry is defined up to a rescaling \cite{Ashtekar:2014zsa}. Using this freedom, in Bondi coordinates $(u,z, \bar z)$, the structure can be taken to be \cite{Ruzziconi:2026bix}
\begin{equation}\label{eq: flatmetric}
    q_{ab} dx^a dx^b = 2 d z d\bar{z}, \qquad n^a \partial_a = \partial_u .
\end{equation} The BMS symmetries, which are the conformal symmetries of this Carrollian structure, satisfy
\begin{equation}
    \mathcal{L}_{\xi}q_{ab}=2\alpha q_{ab} , \qquad \mathcal{L}_{\xi}n^c= -\alpha n^c,
\end{equation}
for a function $\alpha$. The vector fields $\xi$ can be written in terms of the supertranslation generators $\mathcal{T}(z,\bar z)$ and the holomorphic and anti-holomorphic superrotation generators $\mathcal{Y}(z),\mathcal{\bar Y}(\bar z ) $ as 
\begin{equation}\label{eq:zeta}
    \xi=\left(\mathcal{T}+u \alpha \right)\partial_u+\mathcal{Y}\partial_z+\bar{\mathcal{Y}}\partial_{\bar z}, \qquad \alpha=\frac{1}{2}\left(\partial_{z}\mathcal{Y}(z)+\partial_{\bar z}\bar{\mathcal{Y}}(\bar z)\right).
\end{equation}
The global
conformal Carrollian subalgebra is generated by
\begin{equation}
    \mathcal{T}(z,\bar z)=1,z,\bar z,z\bar z;\quad\quad \mathcal{Y}(z)=1,z,z^2;\quad\quad\bar{\mathcal{Y}}(\bar z)=1,\bar z,\bar z^2.
\end{equation}
These 10 generators correspond to the Poincaré generators of the bulk spacetime, that generate translations and Lorentz transformations.

\subsection{Carrollian stress tensor multiplet}

A conformal Carrollian primary multiplet $\Psi_{(h, \bar h)} (u,x^A)$ transforms as \cite{Bagchi:2019xfx, Nguyen:2023vfz, Ruzziconi:2024kzo, Ruzziconi:2026bix}
\begin{equation}  \delta_{\xi}\Psi_{(h,\bar h)}(u,x^A)= \left(
f\partial_u
+ \mathcal{Y} \partial_z+ \bar{\mathcal{Y}} \partial_{\bar z}
+ h\,\partial_z \mathcal{Y}+\bar h\partial_{\bar z}\mathcal{\bar Y}
+ i\partial^B f\mathcal{B}_B\right)\Psi_{(h,\bar h)},
\end{equation} with $x^A = (z, \bar z)$, $f\equiv\mathcal{T}+\frac{u}{2}\left(\partial_z\mathcal{Y}+\partial_{\bar z}\mathcal{\bar Y}\right)$, and $\mathcal{B}_A$ are the Carrollian boost parameters, whose action on the multiplet components
allows for them to mix into each other. Here $h$ and $\bar h$ are the holomorphic and anti-holomorphic weights of the representation, which are related to the spin $s$ and the conformal dimension $\Delta$ as $h=\frac{\Delta+s}{2}, \bar h=\frac{\Delta-s}{2}$. The above primary is called a Carrollian primary singlet if it is irreducible and has $\mathcal{B}_A = 0$ in the above transformation. 

The Carrollian stress tensor ${\mathcal{C}^a}_b$, which satisfies the classical properties\footnote{In principle, these admit an uplift in the quantum theory \cite{Ruzziconi:2024kzo}.} \cite{Ruzziconi:2026bix}
\begin{equation}\label{eq: stresstensorsymmetry}
    {\mathcal{C}^a}_a=0,\quad\quad {\mathcal{C}^A}_B={\mathcal{C}^B}_A,\quad\quad {\mathcal{C}^A}_u=0, \quad\quad\partial_a{\mathcal{C}^a}_b=0, 
\end{equation} is an example of conformal Carrollian multiplet with $\Delta = 3$.\footnote{In this analysis, we assume there is no radiation at null infinity, see e.g. \cite{Donnay:2022aba,Fiorucci:2025twa} for a more general set-up.} Its components are called the Carrollian momenta and we write
\begin{equation}
   \mathcal{C}^u_{\ u} = \mathcal{M},\qquad \mathcal{C}^u_{\ A} = \mathcal{N}_A \, .
\end{equation}
 The relation between the Carrollian momenta and the bulk metric has been discussed in \cite{Donnay:2022aba, Donnay:2022wvx, Fiorucci:2025twa, Hartong:2025jpp}. In particular, $\mathcal{M}$ and $\mathcal{N}_A$ are related to the Bondi mass and angular momentum aspects, respectively. The conformal Carrollian symmetries (\ref{eq:zeta}) act on the Carrollian momenta as:\footnote{The transformations \eqref{eq: stresstensortrafo1}, \eqref{eq: stresstensortrafo2} and \eqref{eq: stresstensortrafo3} are consistent with \cite{Barnich:2013axa,Barnich:2019vzx} for $\mathcal{M}=\text{Re}\,\Psi^0_2$, $\mathcal{N}_{{z}}=\Psi^0_1$, $\mathcal{N}_{\bar z}=\bar\Psi^0_1$ in the absence of radiation, namely when $\Psi^0_4 = 0$, $\Psi^0_3 = 0$ and $\text{Im}\,\Psi^0_2 = 0$; see also \cite{Barnich:2021dta,Barnich:2022bni}. These transformations are consistent with, but not uniquely fixed by the algebra of integrated charges.}  
\begin{align}
\delta_\xi \mathcal{M} &=
\left(
f\,\partial_u + \mathcal{Y} \partial_z+ \bar{\mathcal{Y}} \partial_{\bar z}+ \frac{3}{2}\partial_z \mathcal{Y}+\frac{3}{2} \partial_{\bar z}\mathcal{\bar Y}
\right) \mathcal{M},\label{eq: stresstensortrafo1}\\
    \delta_\xi \mathcal{N}_z &=
\left(
f\,\partial_u + \mathcal{Y} \partial_z+ \bar{\mathcal{Y}} \partial_{\bar z} + 2\partial_z \mathcal{Y}+ \partial_{\bar z}\mathcal{\bar Y}
\right)\mathcal{N}_z
+ 3\partial_z f\, \mathcal{M},\label{eq: stresstensortrafo2} \\
 \delta_\xi \mathcal{N}_{\bar z} &=
\left(
f\,\partial_u + \mathcal{Y} \partial_z+ \bar{\mathcal{Y}} \partial_{\bar z} + \partial_z \mathcal{Y}+2 \partial_{\bar z}\mathcal{\bar Y}
\right)\mathcal{N}_{\bar z}
+ 3\partial_{\bar z} f\, \mathcal{M} . \label{eq: stresstensortrafo3}
\end{align}
These symmetry transformations contain a homogeneous part that coincides with the symmetry transformation of a conformal Carrollian quasi-primary singlet representation, but for $\mathcal{N}_A$  they also contain inhomogeneous terms multiplied by $\partial f$, that display how the components of the stress tensor mix into each other under conformal Carrollian symmetry transformations. 

\subsection{Boundary stress tensor two-point functions}

We have assembled all the necessary ingredients to compute two-point correlators of the Carrollian momenta, by constraining them using the symmetry action displayed in Equations \eqref{eq: stresstensortrafo1}-\eqref{eq: stresstensortrafo3} and imposing the Ward identities 
\begin{equation}
    \delta_\xi\avg{{\C^a}_b{\C^c}_d}=\avg{(\delta_\xi{\C^a}_b){\C^c}_d}+\avg{{\C^a}_b(\delta_\xi{\C^c}_d)}=0.
\end{equation} We sketch the derivation in this section and refer to Appendix \ref{sec: Appendix Boundary correlators} for the detailed computation. 

We start by constraining the two-point function of the rescaled Bondi mass aspect $\avg{\M(u_1,z_1,\bar z_1)\M(u_2,z_2,\bar z_2)}$, which will be shown to vanish, in agreement with the comment below \eqref{matching M with T}. The momentum $\mathcal{M}$ transforms as in \eqref{eq: stresstensortrafo1}, which is identical to the transformation of a primary singlet. Hence, it reproduces the known structure of the Carrollian two-point function for primary singlets \cite{Chen:2021xkw,Donnay:2022wvx}, which is a sum of magnetic (time-independent) and electric (time-dependent) branches:
\begin{equation}\label{eq: Carrollian two-point function}
    \langle\Psi_{(h_1,\bar{h}_1)}(u_1,z_1,\bz_1)\Psi_{(h_2,\bar{h}_2)}(u_2,z_2,\bz_2)\rangle=\frac{C}{z_{12}^{2h_1}\bar z_{12}^{2\bar h_1}}\delta_{h_1,h_2}\delta_{\bar h_1,\bar h_2}+\frac{C'}{u^{h_1+\bar h_1+ h_2+\bar h_2-2}_{12}}\delta_{s_1,-s_2}\delta^{(2)}(z_{12}).
\end{equation} Specifying this for the Carrollian weights of $\mathcal{M}$ yields:
\begin{equation}
    \avg{\M\M}=\frac{C_{\M\M}}{z_{12}^3\bar z_{12}^3}+\frac{C'_{\M\M}}{u_{12}^4}\delta^{(2)}(z_{12}).
\end{equation}
This is not the final result yet, as $\M$ is part of a multiplet and receives additional constraints from the symmetry transformation of correlators $\avg{\M\N_A}$ that mix into $\avg{\mathcal{MM}}$. This mixing appears in the Carrollian boost transformation, which gives the following constraint equation for $\avg{\M\N_z}$ and $\avg{\M\N_{\bar z}}$:
\begin{align}
    &z_{12}\partial_{u_{12}}\avg{\M\mathcal{N}_z}+3\avg{\M\M}= 0 =\bar z_{12}\partial_{u_{12}}\avg{\mathcal{M}\N_z},\label{eq: CBoost1}\\
    &\bar z_{12}\partial_{u_{12}}\avg{\mathcal{M}\mathcal{N}_{\bar z}}+3\avg{\M\M}= 0 = z_{12}\partial_{u_{12}}\avg{\M\mathcal{N}_{\bar z}}.\label{eq: CBoost2}
\end{align}
We can now constrain $\avg{\M\mathcal{N}_z}$ to check whether the above relation is consistent with the other Carrollian symmetry constraints for $\avg{\M\mathcal{N}_z}$.

Let us focus on \eqref{eq: CBoost1}. The only possible solution to these equations occurs for $C_{\mathcal{M}\M} = 0$ and is given by:
\begin{equation}
\avg{\M\N_z} =\frac{C_{\mathcal{M}\mathcal{N}_z}}{z_{12}^\frac{7}{2}\bar z_{12}^{\frac{5}{2}}}+\frac{C'_{\mathcal{M}\mathcal{N}_z}}{u_{12}^3}\delta'(z_{12})\delta(\bar z_{12}),
\end{equation} which gives the identification $C'_{\mathcal{MM}}=-C'_{\mathcal{M}\mathcal{N}_z}$. However, one can check that this correlation function is not compatible with Carrollian special conformal transformation invariance, forcing $\avg{\M\N_z} = 0$ and thus $C'_{\M\M} = 0$. In that case, Equations \eqref{eq: CBoost1} and \eqref{eq: CBoost2} yield the usual constraints on a Carrollian two-point function of Carrollian primary singlets. Since $\langle \mathcal{M} \mathcal{M} \rangle = 0$, this observation is true for all the other constraints, and the correlator $\langle \mathcal{M} \mathcal{N}_A  \rangle$ is of the general form \eqref{eq: Carrollian two-point function}. Because of constraints on the Carrollian weights, this correlator has to vanish, i.e. $\langle \mathcal{M} \mathcal{N}_A  \rangle = 0$.

This implies that the symmetry constraints on the $\langle \mathcal{N}_A \mathcal{N}_B  \rangle$ correlators do not contain any non-zero mixing terms, and thus reduce to the known constraints on two-point functions of Carrollian primary singlets. Therefore, the general solution for these correlators is of the form \eqref{eq: Carrollian two-point function}. More explicitly, taking the specific weights into account, we have
\begin{align}
    \avg{ \mathcal{N}_z \mathcal{N}_{\bar z}}=\frac{C'_{\N_{z}\N_{\bar z}}}{u_{12}^4}\delta^{(2)}(z_{12}),\qquad
    \avg{\N_z\N_z}=\frac{C_{\N_z\N_z}}{z_{12}^{4}\bar z_{12}^2},\qquad
    \avg{\N_{\bar z}\N_{\bar z}}=\frac{C_{\N_{\bar z}\N_{\bar z}}}{z_{12}^{2}\bar z_{12}^4}.
\end{align}
In theories with bulk gravity, these correlators may be further constrained by mixing with other operators in the spectrum. 
It would be interesting to further determine the constants in these expressions, either from a bulk analysis, or by investigating whether they could be related to a central term in the extended BMS algebra (see Appendix 2 of \cite{Barnich:2011ct}). We keep this discussion for future analysis. 

The upshot for us is that, in the vacuum, the two-point function $\langle \mathcal{M} \M \rangle$ vanishes, in agreement with the explicit bulk computation. In the next section, we consider instead insertions of $\mathcal{M}$ in the presence of non-trivial scattering states, and discuss the information on the $\mathcal{S}$-matrix that can be extracted from these insertions. 

\section{In-in correlation functions}\label{sec:inin}

When massive flux passes through $i^\pm$, soft gravitons propagate information about the ingoing or outgoing massive state to the Bondi mass aspect near $\scri^\pm_\pm$, as seen in Section \ref{sec:Massive encoding at I}. As such, measurements of the Bondi mass aspect at $\scri^+_+$ will be related to massive cross sections. In this section, we compute expectation values of the rescaled late-time Bondi mass aspect $\mathcal{M}(\vec{n})$ at $\scri^+_+$ in a given ingoing massive state and relate these correlation functions to averaged massive scattering cross sections. 

\subsection{Background}

In-in correlation functions will compute the expectation value of some detector operator in a given ingoing state. In this formalism, we consider a massive interacting quantum field $\phi$, assuming the following properties \cite{Caron-Huot:2023vxl, Caron-Huot:2023ikn, Moult:2025njc}: 
\begin{itemize}
    \item[a)] The algebra of operators in the asymptotic past is generated by creation and annihilation operators obeying the following commutation relation for operators of the same particle kind: 
    \begin{equation}
        [a_i,a_j^\dagger] = 2p_i^0(2\pi)^3\delta(\vec{p}_i-\vec{p}_j).
    \end{equation}
    \item[b)] An equivalent algebra exists in the asymptotic future, denoted by $b_i, b_j^\dagger$. These are related by unitary evolution $\mathcal{S}$ to the $a_i,a_j^\dagger$ operators:
    \begin{equation}
        b_i = \mathcal{S}^\dagger a_i \mathcal{S},\qquad b_i^\dagger = \mathcal{S}^\dagger a_i^\dagger \mathcal{S}.
    \end{equation}
    \item[c)] There exists a time-independent vacuum state:
    \begin{equation}
        a_i\ket{0} = b_i\ket{0} = 0, \qquad \mathcal{S}\ket{0} = \ket{0}.
    \end{equation}
    \item[d)] Single external particles are stable:
    \begin{equation}
        b_i^\dagger\ket{0} = \mathcal{S}^\dagger a_i^\dagger\ket{0} = a_i^\dagger \ket{0} .
    \end{equation}
\end{itemize} 
In- and out-states are constructed by acting on the vacuum with some number of creation and annihilation operators:
\begin{equation}
    \ket{p_1,\ldots,p_m}_{in} = a^\dagger_{1}\cdots a^\dagger_{m}\ket{0},\ \ket{p_1,\ldots,p_m}_{out} = b^\dagger_{1}\cdots b^\dagger_{m}\ket{0}.
\end{equation}
$\mathcal{S}$-matrix elements are computed as inner product of in- and out-states:
\begin{equation}
\lbraket{out}{in}{p_{n+1},\ldots,p_{m+n}}{p_{1},\ldots,p_n} = \lbra{in}{p_{n+1},\ldots,p_{n+m}}\mathcal{S}\lket{in}{p_1,\ldots,p_n}.
\end{equation}
The in-in correlators of interest will be of the form 
\begin{equation}
   \lbra{\inte}{\psi} \prod_{j \in J}b_j^\dagger \prod_{i \in I}b_i \lket{\inte}{\psi},
\end{equation}
where $\lket{\inte}{\psi}$ is an ingoing state. These can be evaluated by inserting $1 = \sumint_k \lket{\out}{k}\lbra{\out}{k}$:
\begin{equation}
\begin{split}
 \lbra{\inte}{\psi}
  \prod_{j\in J} b_j^\dagger
  \prod_{i\in I}& b_i
  \lket{\inte}{\psi} =
  \sumint_{k}\lbraket{\inte}{\out}{\psi}{k,p_J}\lbraket{\out}{\inte}{k,p_I}{\psi},
  \end{split}
\end{equation}
where $\ket{k,p_J}_{out} = \prod_{j\in J}b_j^\dagger \ket{k}_{out}$ and $\bra{k,p_I}_{out} = \bra{k}\prod_{i\in I}b_i$. In particular, when $I = J$, we can recognize this as a sum over $\mathcal{S}$-matrix elements:
\begin{equation}
\lbra{\inte}{\psi}
  \prod_{j\in J} b_j^\dagger b_j
  \lket{\inte}{\psi}
  =
  \sumint_{k}\,
  |\lbraket{\out}{\inte}{k,p_I}{\psi}|^2.
\end{equation}
\subsection{In-in correlators of $T_{\tau\tau}^{(3)}$}
When acting on the vacuum, the operators $b_j^\dagger$ will create states that look like plane wave states in the asymptotic future. Assuming that at late times, particles spread out and become approximately free, we can write the leading order component of the stress tensor near $i^+$ as 
\begin{equation}
    T^{(3)}_{\tau\tau}(m\hat{x}) = \frac{m^3}{2(2\pi)^3} b(m\hat{x})^\dagger b(m\hat{x}).
\end{equation}
The creation and annihilation operators obey the algebra
\begin{equation}\label{eq: commutator}
    [b(m\hat{x}_1), b^\dagger(m\hat{x}_2)] = 2(2\pi)^3m^{-2}\delta^{(3)}(\hat{x}_1,\hat{x}_2).
\end{equation}
Then, in-in correlators of $T^{(3)}_{\tau\tau}(m\hat{x})$ are computed as sums over cross sections as 
\begin{equation}
    \begin{split}
        \lbra{\inte}{\psi} \prod_{j=1}^n T^{(3)}_{\tau\tau}(m\hat{x}_j)\lket{\inte}{\psi} &= \frac{m^{3n}}{2^n(2\pi)^{3n}}\lbra{\inte}{\psi}\prod_{j=1}^n b(m\hat{x}_j)^\dagger b(m\hat{x}_j)\lket{\inte}{\psi} \\
        &= \frac{m^{3n}}{2^n(2\pi)^{3n}}\sumint_{k} |\lbraket{\out}{\inte}{k,m\hat{x}_1,\ldots,m\hat{x}_n}{\psi}|^2,
    \end{split}
\end{equation}
provided that none of the $T^{(3)}_{\tau\tau}$s are inserted at coincident points; this case will be treated in the next section. The sum is over a complete set of outgoing states of cross sections for the ingoing particles to scatter into, plus additional particles whose creation and annihilation operators appear in the late time limit of the stress tensor. 
\subsection{In-in correlators at $\mathscr{I}^+_+$}
We can now apply this to compute in-in correlators of our massive detector operator $\mathcal{M}$:
\begin{equation}
    \lbra{\inte}{\psi}\prod_{j=1}^n \mathcal{M}(\vec{n}_j)\lket{\inte}{\psi} = (-2)^{-n} \lbra{\inte}{\psi} \int \prod_{j=1}^n d^3\hat{x}_j\sqrt{h_j}G_{Bb}(\hat{x}_j;\vec{n}_j)T^{(3)}_{\tau\tau}(m\hat{x}_j) \lket{\inte}{\psi}.
\end{equation}
If all of the $\hat{x}_j$ are distinct, this is easily written as an integral over sums of squared amplitudes. However, for $n \ge 2$, the integrals will always contain loci where points become coincident, and we need to be careful as these will generate distributional terms that will affect the final result. 

As an illustrative example, consider the case $n = 2$. Then, 
\begin{equation}
    \begin{split}
        \lbra{\inte}{\psi}\mathcal{M}(\vec{n}_1)\mathcal{M}(\vec{n}_2)\lket{\inte}{\psi} = \frac{m^6}{16(2\pi)^6} &\int d^3\hat{x}_1\sqrt{h_1}d^3\hat{x}_2\sqrt{h_2} G_{Bb}(\hat{x}_1;\vec{n}_1)G_{Bb}(\hat{x}_2;\vec{n}_2) \\
        &\times   \lbra{\inte}{\psi} b(m\hat{x}_1)^\dagger b(m\hat{x}_1)b(m\hat{x}_2)^\dagger b(m\hat{x}_2)  \lket{\inte}{\psi}.
    \end{split}
\end{equation}
When $\hat{x}_1 \ne \hat{x}_2$, we can commute all the creation operators to the left and all the annihilation operators to the right. However, because the integral always contains an integral over the locus $\hat{x}_1 = \hat{x}_2$, we have to keep track of this term, obtaining:
\begin{equation}
\begin{split}
       \lbra{\inte}{\psi}\mathcal{M}(\vec{n}_1)\mathcal{M}(\vec{n}_2)\lket{\inte}{\psi} = \frac{m^6}{16(2\pi)^6} &\int d^3\hat{x}_1\sqrt{h_1}d^3\hat{x}_2\sqrt{h_2} G_{Bb}(\hat{x}_1;\vec{n}_1)G_{Bb}(\hat{x}_2;\vec{n}_2) \\
        &\times \lbra{\inte}{\psi}b(m\hat{x}_1)^\dagger b(m\hat{x}_2)^\dagger b(m\hat{x}_1) b(m\hat{x}_2) \lket{\inte}{\psi} \\
        &+ \frac{m^6}{16(2\pi)^6} \int d^3\hat{x}_1\sqrt{h_1}d^3\hat{x}_2\sqrt{h_2}G_{Bb}(\hat{x}_1;\vec{n}_1)G_{Bb}(\hat{x}_2;\vec{n}_2) \\
        &\times \lbra{\inte}{\psi}b(m\hat{x}_1)^\dagger [b(m\hat{x}_1),b(m\hat{x}_2)^\dagger] b(m\hat{x}_2)\lket{\inte}{\psi}.
    \end{split}
\end{equation}

The commutation relation \eqref{eq: commutator} implies that 
\begin{equation}
\begin{split}
       \lbra{\inte}{\psi}\mathcal{M}(\vec{n}_1)\mathcal{M}(\vec{n}_2)\lket{\inte}{\psi}= \frac{m^6}{16(2\pi)^6} &\int d^3\hat{x}_1\sqrt{h_1}d^3\hat{x}_2\sqrt{h_2}G_{Bb}(\hat{x}_1;\vec{n}_1)G_{Bb}(\hat{x}_2;\vec{n}_2) \\
        &\times \lbra{\inte}{\psi}b(m\hat{x}_1)^\dagger b(m\hat{x}_2)^\dagger b(m\hat{x}_1) b(m\hat{x}_2)\lket{\inte}{\psi} \\
        &+ \frac{m^4}{8(2\pi)^3} \int d^3\hat{x}_1\sqrt{h_1}G_{Bb}(\hat{x}_1;\vec{n}_1)G_{Bb}(\hat{x}_1;\vec{n}_2) \\
        &\times \lbra{\inte}{\psi}b(m\hat{x}_1)^\dagger b(m\hat{x}_1)\lket{\inte}{\psi}.
    \end{split}
\end{equation}

Because each of the terms now has creation operators always to the left of annihilation operators, we can now relate the two-point stress-tensor in-in correlator at late times to sums over squares of scattering amplitudes:
\begin{equation}
    \begin{split}
        \lbra{\inte}{\psi}\mathcal{M}(\vec{n}_1)\mathcal{M}(\vec{n}_2)\lket{\inte}{\psi} &= \frac{m^6}{16(2\pi)^6}\int d^3\hat{x}_1\sqrt{h_1}d^3\hat{x}_2\sqrt{h_2} G_{Bb}(\hat{x}_1;\vec{n}_1)G_{Bb}(\hat{x}_2;\vec{n}_2) \\
        &\times \sumint_{k^{out}} |\lbraket{\out}{\inte}{k^{out}, m\hat{x}_1, m\hat{x}_2}{\psi}|^2 \\
        &+ \frac{m^4}{8(2\pi)^3}\int d^3\hat{x}\sqrt{h} G_{Bb}(\hat{x};\vec{n}_1)G_{Bb}(\hat{x};\vec{n}_2) \\
        &\times \sumint_{k^{out}} |\lbraket{\out}{\inte}{k^{out}, m\hat{x}}{\psi}|^2.
    \end{split}
\end{equation}
We can extend this iteratively to derive a diagrammatic expansion of the $n$-point stress-tensor in-in correlation function. 

Diagrams for computing $\lbra{\inte}{\psi}\mathcal{M}(\vec{n})^n\lket{\inte}{\psi}$ are constructed iteratively as follows. 
\begin{enumerate}
    \item Start with $n$ marked points on the boundary of a circle. Draw a line connecting each point to a different point in the interior of the circle. This diagram represents the contribution where each interior $T^{(3)}_{\tau\tau}(m\hat{x}_i)$ is inserted at a different point and corresponds to 
    \begin{equation}
        \begin{split}
            \frac{(-1)^nm^{3n}}{4^n(2\pi)^{3n}}\int \prod_{i=1}^n d^3\hat{x}_i\sqrt{h_i} G_{Bb}(\hat{x}_i;\vec{n}_i) \sumint_{k^{out}} |\lbraket{\out}{\inte}{\{k^{out}, m\hat{x}_1,\ldots,m\hat{x}_n\}}{\psi}|^2.
        \end{split}
    \end{equation}
    \item Starting with a valid diagram as above, a new diagram can be generated by putting two interior points $\hat{x}_i$ and $\hat{x}_j$ on the same point. The contribution attributed to this diagram is obtained from that of the parent diagram by:
    \begin{enumerate}
        \item Multiplying by $2(2\pi)^3m^{-2}$,
        \item Setting $\hat{x}_j = \hat{x}_i$ and dropping the integral over $\hat{x}_j$,
        \item and removing the $j$-th outgoing particle. 
    \end{enumerate}
\end{enumerate}
The final answer will be the sum over all diagrams constructed in this manner. Note that the same diagram could be generated in several different methods with this expansion; such diagrams should only be included once. Several of these diagrams for the case $n = 5$ are shown in Figure \ref{fig:diagexp} below:
\begin{figure}[h!]
\begin{center}
\begin{tikzpicture}[scale=1.2,
  outerpt/.style={circle, fill, inner sep=1.4pt},
  innerpt/.style={circle, fill, inner sep=1.3pt},
  conn/.style={line width=0.7pt}
]

\newcommand{\FivePointCircle}[4]{%
  \draw (#2,#3) circle (#4);
  \foreach \i in {1,...,5} {%
    \coordinate (#1P\i) at ($(#2,#3) + ({90+72*(\i-1)}:#4)$);
    \node[outerpt] at (#1P\i) {};
  }%
}

\def\R{1.2}
\FivePointCircle{A}{0}{0}{\R}
\FivePointCircle{B}{3.2}{0}{\R}
\FivePointCircle{C}{6.4}{0}{\R}
\FivePointCircle{D}{9.6}{0}{\R}

\foreach \i in {1,...,5} {
  \coordinate (AI\i) at ($(0,0) + ({90+72*(\i-1)+36}:0.55)$);
  \node[innerpt] at (AI\i) {};
  \draw[conn] (AP\i) -- (AI\i);
}

\coordinate (BI1) at (3.2,0.55);   \node[innerpt] at (BI1) {};
\coordinate (BI2) at (2.75,0.1);  \node[innerpt] at (BI2) {};
\coordinate (BI3) at (3.65,-0.10); \node[innerpt] at (BI3) {};
\coordinate (BI4) at (3.15,-1); \node[innerpt] at (BI4) {};

\draw[conn] (BP1) -- (BI1);
\draw[conn] (BP2) -- (BI1);  
\draw[conn] (BP3) -- (BI3);
\draw[conn] (BP4) -- (BI4);
\draw[conn] (BP5) -- (BI2);

\coordinate (CI1) at (6.4,0.45);   \node[innerpt] at (CI1) {};
\coordinate (CI2) at (6.4,-0.45);  \node[innerpt] at (CI2) {};

\draw[conn] (CP1) -- (CI1);
\draw[conn] (CP2) -- (CI1);
\draw[conn] (CP3) -- (CI1);

\draw[conn] (CP4) -- (CI2);
\draw[conn] (CP5) -- (CI2);

\coordinate (DI1) at (9.6,0); \node[innerpt] at (DI1) {};
\foreach \i in {1,...,5} {
  \draw[conn] (DP\i) -- (DI1);
}


\end{tikzpicture}
\end{center}
\caption{A sample of diagrams associated to computing $\lbra{\inte}{\psi}(\mathcal{M})^5\lket{\inte}{\psi}$. Diagrams with multiple external points connected to the same internal point are generated by commuting $b$s and $b^\dagger$s past each other. \label{fig:diagexp}} 
\end{figure}
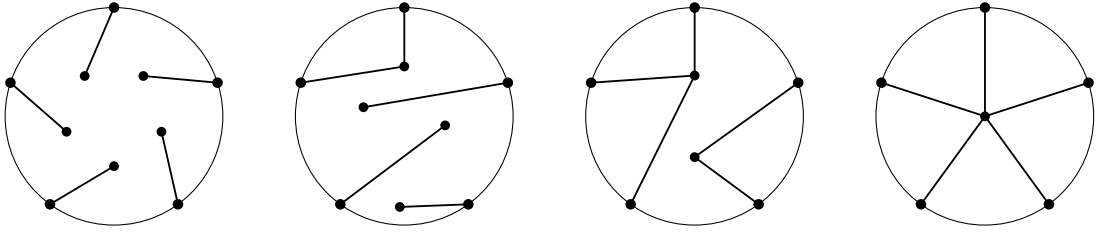
\section{Applications}\label{sec:apps}
As seen above, in-in correlators of the detector operator $\mathcal{M}$  can be computed as weighted sums over $S$-matrix elements. In general, however, these functions are unwieldy both because of the large number of diagrams above to sum over and because we cannot discard disconnected contributions to the $S$-matrix due to the integration over the direction of all outgoing particles. However, when the ingoing state is either a one- or two-particle state, there are only a small set of amplitudes that can contribute at low order in the appropriate massive coupling constant. 

\subsection{Ingoing state reconstruction for free massive scalars}
First, we consider the case where the ingoing state is a one-particle wavepacket 
\begin{equation}
    \lket{\inte}{\psi} = \int \widetilde{d^3p}\psi(p)a(p)^\dagger\ket{0},
\end{equation}
where $\widetilde{d^3p}$ is the Lorentz invariant measure defined as $\widetilde{d^3p}=\frac{d^3p}{(2\pi)^32p^0}=\frac{m^2d^3\hat{x} \sqrt{h}}{2(2\pi)^3}$. When evaluating the correlation function 
\begin{equation}
    \lbra{\inte}{\psi}\prod_{j=1}^n \mathcal{M}(\vec{n}_j)\lket{\inte}{\psi},
\end{equation}
the only contribution we can get must arise from integrating over a single-outgoing particle, as any $1 \to n$ amplitude vanishes kinematically for $n>1$, and we have that 
\begin{equation}
    \begin{split}
        \lbra{\inte}{\psi}\prod_{j=1}^n \mathcal{M}(\vec{n}_j)\lket{\inte}{\psi} &= \frac{(-1)^nm^{n+2}}{2^{4+n}\pi^3}\int d^3\hat{x} \sqrt{h}\prod_{j=1}^n G_{Bb}(\hat{x};\vec{n}_j)  |\psi(m\hat{x})|^2.
    \end{split}
\end{equation}

We can extract the energy-weighted norm of the massive bulk particle from the stress tensor inserted correlator. We consider the one-point insertion:
\begin{equation}
     \lbra{\inte}{\psi}\mathcal{M}(\vec{n})\lket{\inte}{\psi} = -\frac{m^{3}}{2^{5}\pi^3}\int d^3\hat{x} \sqrt{h} G_{Bb}(\hat{x};\vec{n})  |\psi(m\hat{x})|^2
\end{equation}
and use the known integral of the Green's function over the sphere \cite{Campiglia:2015lxa}:
\begin{equation}
   \int d^2\vec{n}G^{}_{Bb}(\hat{x};\vec{n})=-2\cosh{\rho}=-2\frac{p^0}{m}
\end{equation}
to find the weighted integrated norm as: 
\begin{equation}
\begin{split}
    \int d^3\hat{x} \sqrt{h}{p^0}|\psi(m\hat{x})|^2&=\frac{2(2\pi)^3}{m^2}\int d^2\vec{n}\lbra{\inte}{\psi}\mathcal{M}(\vec{n})\lket{\inte}{\psi},
\end{split}
\end{equation}
where we used Eq. \eqref{eq: momentum}. This can be rewritten in momentum space:
\begin{equation}
    \int \widetilde{d^3p}\, p^0|\psi({p})|^2=\int d^2\vec{n}\lbra{\inte}{\psi}\mathcal{M}(\vec{n})\lket{\inte}{\psi}.
\end{equation}
\subsection{Two-particle scattering}

The above formulas also simplify when considering a state with only two ingoing particles. In this case, in-in correlation functions can be evaluated as a sum over $S$-matrix elements with two ingoing and some number $n_{\out}$ of outgoing particles. The case $n_{\out} = 0, 1$ are forbidden kinematically, and for $n_{\out} \ge 3$ the only contribution comes from fully connected amplitudes as any disconnected contribution would include a $1 \to n$ amplitude with $n > 1$, which is again forbidden kinematically. Additionally, at small coupling (which we hereafter assume) these will be subleading to the $n_{out} = 2$ case. 

Consider now the $n$-point in-in correlation function $\lbra{\inte}{\psi}\prod_{j=1}^n \mathcal{M}(\vec{n}_j)\lket{\inte}{\psi}$ where $\lket{\inte}{\psi}$ is a general two-particle ingoing state $\lket{\inte}{\psi} = \int \widetilde{d^3p_1}\widetilde{d^3p_2}\psi(p_1,p_2)a_1^\dagger a_2^\dagger\ket{0}$. The leading contributions to this will arise from $2 \to 2$ scattering processes, as contributions from higher point amplitudes will necessarily be connected and subleading at small coupling. 

In this state, the one-point in-in correlator of $\mathcal{M}$ takes the form 
\begin{equation}
    \begin{split}
        \lbra{\inte}{\psi}\mathcal{M}(\vec{n})\lket{\inte}{\psi} &= - \frac{m^3}{32\pi^3}\int d^3\hat{x}\sqrt{h}G_{Bb}(\hat{x};\vec{n})\int\widetilde{d^3k}|\lbraket{\out}{\inte}{k,m\hat{x}}{\psi}|^2 \\
        &= -\frac{m^3}{8\pi^3}\int d^3\hat{x}\sqrt{h}G_{Bb}(\hat{x};\vec{n})\int \widetilde{d^3k} |\psi(m\hat{x},k)|^2 \\
        &+\frac{m^3}{(2\pi)^2}\int d^3\hat{x}\sqrt{h}G_{Bb}(\hat{x};\vec{n})\int\widetilde{d^3k} \\
        &\times \mathrm{Im}\left[\int \widetilde{d^3p_1}\delta_+((m\hat{x}+k-p_1)^2+m^2) \psi^*(m\hat{x},k)\psi(p_1,m\hat{x}+k-p_1) M_{p_1,m\hat{x}+k-p_1\to m\hat{x}k}\right] \\
        &-8\pi^5 m^3\int d^3\hat{x}\sqrt{h}G_{Bb}(\hat{x};\vec{n})\int \widetilde{d^3k}\prod_{j=1}^4 \widetilde{d^3p_j}\psi^*(p_3,p_4)\psi(p_1,p_2) \\
        &M^\dagger_{p_3p_4\to m\hat{x}k}M_{p_1p_2\to m\hat{x},k}\delta^{(4)}(p_3+p_4-p_1-p_2)\delta^{(4)}(m\hat{x}+k-p_1-p_2).
    \end{split}
\end{equation}
up to corrections subleading in the coupling. Here we have defined $\delta_+(p^2+m^2) = \delta(p^2+m^2)\Theta(p^0)$. The two-point function takes the form
\begin{equation}
\begin{split}
    \lbra{\inte}{\psi}\mathcal{M}(\vec{n}_1)\mathcal{M}(\vec{n}_2)\lket{\inte}{\psi} &= F_1^2 + F_2^2 \\
    F_1^2 &= \frac{m^4}{16\pi^3}\int d^3\hat{x} \sqrt{h}G_{Bb}(\hat{x};\vec{n}_1)G_{Bb}(\hat{x};\vec{n}_2)\int \widetilde{d^3k}|\psi(m\hat{x},k)|^2 \\
        &-\frac{m^4}{2(2\pi)^2}\int d^3\hat{x}\sqrt{h}G_{Bb}(\hat{x};\vec{n}_1)G_{Bb}(\hat{x};\vec{n}_2)\int\widetilde{d^3k} \\
        &\times \,\mathrm{Im}\bigg[\int \widetilde{d^3p_1}\delta_+((m\hat{x}+k-p_1)^2+m^2)\\
        &\times\psi^*(m\hat{x},k)\psi(p_1,m\hat{x}+k-p_1) M_{p_1,m\hat{x}+k-p_1\to m\hat{x}k}\bigg] \\
        &+ \frac{m^4\pi^2}{2}\int d^3\hat{x}\sqrt{h}G_{Bb}(\hat{x};\vec{n}_1)G_{Bb}(\hat{x};\vec{n}_2) \\
        &\times \int \prod_{j=1}^4 \widetilde{d^3p_j}\psi^*(p_3,p_4)\psi(p_1,p_2)\delta_+((p_1+p_2-m\hat{x})^2+m^2) \\
        &\times M^\dagger_{p_3p_4\to m\hat{x},p_1+p_2-m\hat{x}}M_{p_1p_2\to m\hat{x},p_1+p_2-m\hat{x}}\delta^{(4)}(p_3+p_4-p_1-p_2) \\
        F_2^2 &= \frac{m^6}{2^{10}\pi^6}\int d^3\hat{x}_1\sqrt{h_1}d^3\hat{x}_2\sqrt{h_2}G_{Bb}(\hat{x}_1;\vec{n}_1)G_{Bb}(\hat{x}_2;\vec{n}_2) |\lbraket{\out}{\inte}{m\hat{x}_1,m\hat{x}_2}{\psi}|^2 \\
        &= \frac{m^6}{256\pi^6}\int d^3\hat{x}_1\sqrt{h_1}d^3\hat{x}_2\sqrt{h_2}G_{Bb}(\hat{x}_1;\vec{n}_1)G_{Bb}(\hat{x}_2;\vec{n}_2)|\psi(m\hat{x}_1,m\hat{x}_2)|^2 \\
        &-\frac{m^6}{128\pi^5}\int d^3\hat{x}_1\sqrt{h_1}d^3\hat{x}_2\sqrt{h_2}G_{Bb}(\hat{x}_1;\vec{n}_1)G_{Bb}(\hat{x}_2;\vec{n}_2) \\
        &\times \mathrm{Im}\left[\int \widetilde{d^3p_1}\delta_+((m\hat{x}_1+m\hat{x}_2-p_1)^2+m^2) \psi^*(m\hat{x}_1,m\hat{x}_2)\psi(p_1,m(\hat{x}_1+\hat{x}_2)-p_1) \right. \\
        &\left.\times M_{p_1,m(\hat{x}_1+\hat{x}_2)-p_1\to m\hat{x}_1m\hat{x}_2}\right] \\
        & +\frac{\pi^2m^6}{4}\int d^3\hat{x}_1\sqrt{h_1}d^3\hat{x}_2\sqrt{h_2}G_{Bb}(\hat{x}_1;\vec{n}_1)G_{Bb}(\hat{x}_2;\vec{n}_2) \\
        &\times \int\prod_{j=1}^4 \widetilde{d^3p_j}\psi^*(p_3,p_4)\psi(p_1,p_2)M^\dagger_{p_3p_4\to m\hat{x}_1m\hat{x}_2}M_{p_1p_2\to m\hat{x}_1m\hat{x}_2} \\
        &\times \delta^{(4)}(p_3+p_4-p_1-p_2)\delta^{(4)}(m\hat{x}_1+m\hat{x}_2-p_1-p_2).
\end{split}
\end{equation}
The details of this calculation appear in Appendix \ref{app:twopoint}. 
\section{Conclusion}

All massive trajectories in asymptotically flat spacetimes start at $i^-$ and end at $i^+$. Consequently, in a massive theory on a fixed background, experiments at null infinity will never learn about the massive $S$-matrix. When gravity becomes dynamical, however, soft graviton radiation inevitably carries information about the outgoing massive state at $i^+$ to null infinity, allowing us to define a detector operator for massive outgoing particles. Unlike detector operators for massless particles or the massive detector operator proposed in \cite{Have:2024dff}, such as the energy/ANEC operator, the late time limit of the Bondi mass aspect is sensitive to outgoing massive radiation in every direction. However, it is useful because it can naturally be measured at null infinity. 

Nevertheless, there is more work to be done. While our work suggests that much of the massive $S$-matrix can be extracted from $\mathscr{I}$ given careful measurements, it does not provide a holographic dual for massive scattering because the ingoing bulk state is not prepared holographically. Interestingly, our detector operator $\mathcal{M}$ coincides exactly with the late time limit of the Carrollian stress tensor $\mathcal{C}^u_{\ u}(u,\vec{n})$. It is tempting to speculate that, for a given ingoing state $\lket{\inte}{\psi}$, there could be a state in the Carrollian CFT $\ket{X}$ such that 
\begin{equation}
    \lbra{\inte}{\psi}\prod_{j=1}^n\mathcal{M}(\vec{n}_j)\lket{\inte}{\psi} = \bra{X}\prod_{j=1}^n \lim_{u\to \infty}\mathcal{C}^{u}_{\ u}(u,\vec{n}_j)\ket{X},
\end{equation}
although it remains unclear how $\ket{X}$ could naturally be organized into representations of the conformal Carrollian algebra. It may also be necessary to couple the Carrollian theory to a dual theory on $i^\pm$ to represent ingoing and outgoing massive states; such a dual has been suggested in \cite{Have:2024dff}. 

It would also be interesting to include backreaction in our calculation. Throughout our calculation we have assumed that we have a fixed massive scattering process and are considering the first correction to the metric. While this may in some cases be a reasonable approximation, weak gravitational effects over long time spans can lead to logarithmic deviations to particle trajectories at late times that may modify our results \cite{Choi:2026cyh}. 

\section{Acknowledgements}
The authors would like to thank Tim Adamo, Simon Heuveline, Sruthi Narayanan, Enrico Pajer, Abhi Peringara Sajeev, Jakob Salzer, Volker Schomerus, and Atul Sharma for useful conversations. WM is supported by a Junior Fellowship from the Society of Fellows of Harvard University. KM is supported by the German National Academic Foundation (Studienstiftung des deutschen Volkes) and the MLP Studies Scholarship. RR is supported by the European Union’s Horizon Europe research and innovation programme under the Marie Sklodowska-Curie grant agreement No. 101104845 (UniFlatHolo), hosted at Harvard University and Ecole Polytechnique. Preprint number: CPHT-RR022.082026

\appendix
\section{Details on the derivation of the boundary stress tensor correlators}\label{sec: Appendix Boundary correlators}

Translation symmetry forces $\avg{\M\mathcal{N}_z}$ to be a function of difference variables $(u_{12},z_{12},\bar z_{12})$. The homogeneous part of the Carrollian boost symmetry generated by $\mathcal{T}=z,\bar z$  implies:
\begin{equation}
    \avg{\M\mathcal{N}_z}(u_{12},z_{12},\bar z_{12})=h(z_{12},\bar z_{12})+j(u_{12},z_{12})\delta(\bar z_{12}).
\end{equation}
Then the dilatation symmetry generated by $u_{12} \partial_{u_{12}} + z_{12} \partial_{z_{12}} + \bar{z}_{12} \partial_{\bar z_{12}}$ implies:
\begin{align}
    h(z_{12},\bar z_{12})&=\frac{C_{\mathcal{M}\mathcal{N}_z}}{z_{12}^\rho \bar z_{12}^\sigma}\quad \text{with} \quad \rho+\sigma=6,\\
    j_a(u_{12},z_{12})&=\frac{C'_{\mathcal{M}\mathcal{N}_z}}{u_{12}^{5-a}}\partial_{z_{12}}^{a-1}\delta(z_{12}),
\end{align}
where $a$ is the negative eigenvalue defined by:  $z_{12}\partial_{z_{12}}j_a=-aj_a$. Imposing invariance under spatial rotation gives:
\begin{align}
    \rho-\sigma=1 &\implies\rho=\frac{7}{2},\sigma=\frac{5}{2},\\
    1+s_2=a&\implies a=2.
\end{align}
This allows for writing $\avg{\M\mathcal{N}_z}(u_{12},z_{12},\bar z_{12})$ as:
\begin{equation}\label{eq: MN Correlator}
    \avg{\M\mathcal{N}_z}(u_{12},z_{12},\bar z_{12})=\frac{C_{\mathcal{M}\mathcal{N}_z}}{z_{12}^\frac{7}{2}\bar z_{12}^{\frac{5}{2}}}+\frac{C'_{\mathcal{M}\mathcal{N}_z}}{u_{12}^3}\delta'(z_{12})\delta(\bar z_{12}).
\end{equation}
This can now be plugged into eq.(\ref{eq: CBoost1}) to check whether this is consistent with the mixing in the Carrollian boost transformation. The resulting equation takes the form:
\begin{equation}
    0=\frac{C'_{\mathcal{M}\mathcal{N}_z}+C'_{\mathcal{MM}}}{u_{12}^4}\delta^{(2)}(z_{12})+\frac{C_\mathcal{MM}}{z_{12}^3\bar z_{12}^3},
\end{equation}
which forces $C_\mathcal{MM}=0$ and $C'_{\mathcal{M}\mathcal{N}_z}=-C'_{\mathcal{MM}}$.  

By imposing invariance under ($\mathcal{T}=0,\mathcal{Y}=z^2,\bar{\mathcal{Y}}=0$) and setting $u_2,z_2,\bar z_2=0$, we find the following constraint equation for $\avg{\M\mathcal{N}_z}(u_1,z_1,\bar z_1)$:
\begin{equation}(u_1z_1\partial_{u_1}+z_1^2\partial_{z_1}+3z_1)\avg{\M\mathcal{N}_z}(u_{1},z_{1},\bar z_{1})=0,
\end{equation}
which forces $C_{\mathcal{M}\mathcal{N}_z}=0$. Applying this to the electric branch of $\avg{\M\mathcal{N}_z}$ one finds:
\begin{equation}
    -\frac{2C'_{\mathcal{MM}}}{u_1^3}\delta^{(2)}(z_{1})=0,
\end{equation}
from which we can conclude that $C'_{\mathcal{MM}}=C'_{\mathcal{M}\mathcal{N}_z}=0$ and $\avg{\mathcal{MM}}$ vanishes completely. Since $\avg{\M\N_A}$ only mixes into $\avg{\M\M}$, it must transform only homogeneously as the Carrollian two-point function \eqref{eq: Carrollian two-point function}, which does not allow for non-zero terms for $\avg{\M\N_A}$ because of the constraints on the weights and spins.

\section{In-In Correlators for Two-Particle Scattering}\label{app:twopoint}
In this section, we calculate the one- and two-point in-in correlation functions of $\mathcal{M}$ in a two-particle ingoing state. The $n$-point in-in correlation functions are dominated by terms with the minimum number of outgoing particles, which contribute as 
\begin{equation}
    \begin{split}
        F^n_1(\{\vec{n}_j\}_{j\in \{1,\dots,n\}},\psi) &= \frac{(-1)^nm^{n+2}}{2^{4+n}\pi^3}\int d^3\hat{x} \sqrt{h}\prod_{j=1}^n G_{Bb}(\hat{x},\vec{n}_j) \int \widetilde{d^3k}|\lbraket{\out}{\inte}{k, m\hat{x}}{\psi}|^2, \\
        F^n_2(\{\vec{n}_j\}_{j\in\{1,\dots,n\}},\psi) &= \frac{(-1)^nm^{n+4}}{2^{9+n}\pi^6}\sum_{S \subset \{1,\ldots, n\}} \int d^3\hat{x}_1\sqrt{h_1}d^3\hat{x}_2 \sqrt{h_2} \prod_{i \in S}G_{Bb}(\hat{x}_1;\vec{n}_i)\\&\times\prod_{j \not\in S}G_{Bb}(\hat{x}_2;\vec{n}_j) |\lbraket{\out}{\inte}{m\hat{x}_1,m\hat{x}_2}{\psi}|^2,
    \end{split}
\end{equation}
where the sum in the last line is taken over distinct subsets of $\{1,\ldots, n\}$ and an extra factor of $1/2$ is included to account for summing over both $S$ and $\{1,\ldots,n\}/S$. Other contributions, arising from diagrams with more unconnected internal points or from summing over unobserved external states, will be subleading to these contributions. 

Now, 
\begin{equation}
\begin{split}\label{eq: square}
    |\lbraket{\out}{\inte}{k_1,k_2}{\psi}|^2 &= \lbraket{\inte}{\out}{\psi}{k_1,k_2}\lbraket{\out}{\inte}{k_1,k_2}{\psi} \\
    &= \int \prod_{j=1}^4 \widetilde{d^3p_j} \psi(p_3,p_4)^*\psi(p_1,p_2)\bra{0}a(p_3)a(p_4)b(k_1)^\dagger b(k_2)^\dagger\ket{0}\bra{0}b(k_1)b(k_2)a(p_1)^\dagger a(p_2)^\dagger \ket{0} \\
    &= \int \prod_{j=1}^4 \widetilde{d^3p_j} \psi(p_3,p_4)^*\psi(p_1,p_2)\bra{0}a(p_3)a(p_4)\mathcal{S}^\dagger a(k_1)^\dagger a(k_2)^\dagger\ket{0} \\
    &\times \bra{0}a(k_1)a(k_2)\mathcal{S}a(p_1)^\dagger a(p_2)^\dagger \ket{0}.
\end{split}
\end{equation}
The $2\to 2$ scattering matrix can be expanded as a disconnected piece and the connected amplitude 
\begin{equation}
    \mathcal{S} = 1 + i\mathcal{T}
\end{equation}
where $\mathcal{T}$ describes the connected scattering amplitude. Then, letting $\hat{\delta}(k_1,k_2) = 2(2\pi)^3k_1^0\delta^{(3)}(\vec{k}_{1}-\vec{k}_2)$,
\begin{equation}
    \begin{split}
        \bra{0}a(k_1)a(k_2)\mathcal{S}a(p_1)^\dagger a(p_2)^\dagger\ket{0} &= \bra{0}a(k_1)a(k_2)a(p_1)^\dagger a(p_2)^\dagger\ket{0} + i\bra{0}a(k_1)a(k_2)\mathcal{T}a(p_1)^\dagger a(p_2)^\dagger\ket{0} \\
        &= (\hat{\delta}(k_1,p_1)\hat{\delta}(k_2,p_2) + \hat{\delta}(k_1,p_2)\hat{\delta}(k_2,p_1)) \\
        &+ (2\pi)^4i M_{p_1p_2\to k_1k_2}\delta^{(4)}(p_1+p_2-k_1-k_2).
    \end{split}
\end{equation}
Inserting this into Equation \eqref{eq: square} gives: 
\begin{equation}
    \begin{split}
        |\lbraket{\out}{\inte}{k_1,k_2}{\psi}|^2 
        &= \int \prod_{j=1}^4\widetilde{d^3p_j}\psi_{34}^*\psi_{12} \\
        &\times (\hat{\delta}(p_3,k_1)\hat{\delta}(p_4,k_2) + \hat{\delta}(p_3,k_2)\hat{\delta}(p_4,k_1) - (2\pi)^4iM_{p_3p_4\to k_1k_2}^\dagger \delta^{(4)}(p_3+p_4-k_1-k_2)) \\
        &\times (\hat{\delta}(k_1,p_1)\hat{\delta}(k_2,p_2) + \hat{\delta}(k_1,p_2)\hat{\delta}(k_2,p_1) + (2\pi)^4iM_{p_1p_2\to k_1k_2}\delta^{(4)}(p_1+p_2-k_1-k_2)) \\
        &= \int \prod_{j=1}^4\widetilde{d^3p_j}\psi_{34}^*\psi_{12}(\hat{\delta}(p_3,k_1)\hat{\delta}(p_4,k_2) + \hat{\delta}(p_3,k_2)\hat{\delta}(p_4,k_1)) \\
        &\times (\hat{\delta}(k_1,p_1)\hat{\delta}(k_2,p_2) + \hat{\delta}(k_1,p_2)\hat{\delta}(k_2,p_1)) \\
        &- (2\pi)^4i \int \prod_{j=1}^4\widetilde{d^3p_j}\psi_{34}^*\psi_{12}(\hat{\delta}(k_1,p_1)\hat{\delta}(k_2,p_2) + \hat{\delta}(k_1,p_2)\hat{\delta}(k_2,p_1)) \\
        &\times M_{p_3p_4\to k_1k_2}^\dagger \delta^{(4)}(p_3+p_4-k_1-k_2) \\
        &+ (2\pi)^4i\int \prod_{j=1}^4\widetilde{d^3p_j}\psi_{34}^*\psi_{12}(\hat{\delta}(p_3,k_1)\hat{\delta}(p_4,k_2) + \hat{\delta}(p_3,k_2)\hat{\delta}(p_4,k_1)) \\
        &\times M_{p_1p_2\to k_1k_2}\delta^{(4)}(p_1+p_2-k_1-k_2) \\
        &+ (2\pi)^8\int \prod_{j=1}^4\widetilde{d^3p_j}\psi_{34}^*\psi_{12} M^\dagger_{p_3p_4\to k_1k_2} M_{p_1p_2\to k_1k_2} \\
        &\times\delta^{(4)}(k_1+k_2-p_3-p_4)\delta^{(4)}(k_1+k_2-p_1-p_2).
    \end{split}
\end{equation}
Using the fact that $\psi_{ij} = \psi(p_i,p_j) = \psi_{ji}$, the first term evaluates to 
\begin{equation}
    4|\psi(k_1,k_2)|^2.
\end{equation}
The second becomes:
\begin{equation}
\begin{split}
    -2(2\pi)^4i&\int \widetilde{d^3p_3}\widetilde{d^3p_4} \psi^*(p_3,p_4)\psi(k_1,k_2) M^\dagger_{p_3p_4\to k_1k_2}\delta^{(4)}(p_3+p_4-k_1-k_2). \\
    &= -4\pi i\int\widetilde{d^3p_3}\int d^4p_4\delta(p_4^2+m^2)\Theta(p_4^0)\psi^*_{34}\psi_{12}M^\dagger_{p_3p_4\to p_1p_2}\delta^{(4)}(p_3+p_4-k_1-k_2) \\
    &= -4\pi i\int \widetilde{d^3p_3}\delta_+((-p_3+k_1+k_2)^2+m^2)\psi^*(p_3,k_1+k_2-p_3)\psi(k_1,k_2)M^\dagger_{p_3,k_1+k_2-p_3\to k_1k_2}.
    \end{split}
\end{equation}

The third term is:
\begin{equation}
\begin{split}
    2(2\pi)^4i&\int \widetilde{d^3p_1}\widetilde{d^3p_2} \psi^*(k_1,k_2)\psi(p_1,p_2) M_{p_1p_2\to k_1k_2}\delta^{(4)}(p_1+p_2-k_1-k_2) \\
    &= 4\pi i\int \widetilde{d^3p_1}\delta_+((-p_1+k_1+k_2)^2+m^2)\psi^*(k_1,k_2)\psi(p_1,k_1+k_2-p_1)M_{p_1,k_1+k_2-p_1\to k_1k_2}.
\end{split}
\end{equation}
The final term becomes:
\begin{equation}
    \begin{split}
        (2\pi)^8\int \prod_{j=1}^4\widetilde{d^3p_j}\psi^*(p_3,p_4)\psi(p_1,p_2) M^\dagger_{p_3p_4\to k_1k_2}M_{p_1p_2\to k_1k_2}\delta^{(4)}(p_3+p_4-p_1-p_2)\delta^{(4)}(k_1+k_2-p_1-p_2).
    \end{split}
\end{equation}
We can now use these to evaluate the above correlation function to leading power in the coupling. For $n = 1$, only the first term is present, leading to
\begin{equation}
    \begin{split}
        \lbra{\inte}{\psi}\mathcal{M}(\vec{n})\lket{\inte}{\psi} &= - \frac{m^3}{32\pi^3}\int d^3\hat{x}\sqrt{h}G_{Bb}(\hat{x};\vec{n})\int\widetilde{d^3k}|\lbraket{\out}{\inte}{k,m\hat{x}}{\psi}|^2 \\
        &= -\frac{m^3}{8\pi^3}\int d^3\hat{x}\sqrt{h}G_{Bb}(\hat{x};\vec{n})\int \widetilde{d^3k} |\psi(m\hat{x},k)|^2 \\
        &+\frac{m^3}{(2\pi)^2}\int d^3\hat{x}\sqrt{h}G_{Bb}(\hat{x};\vec{n})\int\widetilde{d^3k} \\
        &\times \mathrm{Im}\left[\int \widetilde{d^3p_1}\delta_+((m\hat{x}+k-p_1)^2+m^2) \psi^*(m\hat{x},k)\psi(p_1,m\hat{x}+k-p_1) M_{p_1,m\hat{x}+k-p_1\to m\hat{x}k}\right] \\
        &-8\pi^5 m^3\int d^3\hat{x}\sqrt{h}G_{Bb}(\hat{x};\vec{n})\int \widetilde{d^3k}\prod_{j=1}^4 \widetilde{d^3p_j}\psi^*(p_3,p_4)\psi(p_1,p_2) \\
        &M^\dagger_{p_3p_4\to m\hat{x}k}M_{p_1p_2\to m\hat{x},k}\delta^{(4)}(p_3+p_4-p_1-p_2)\delta^{(4)}(m\hat{x}+k-p_1-p_2).
    \end{split}
\end{equation}
We can similarly apply this to the case $n = 2$ to obtain:
\begin{equation}
\begin{split}
    \lbra{\inte}{\psi}\mathcal{M}(\vec{n}_1)\mathcal{M}(\vec{n}_2)\lket{\inte}{\psi} &= F_1^2 + F_2^2 \\
    F_1^2 &= \frac{m^4}{16\pi^3}\int d^3\hat{x} \sqrt{h}G_{Bb}(\hat{x};\vec{n}_1)G_{Bb}(\hat{x};\vec{n}_2)\int \widetilde{d^3k}|\psi(m\hat{x},k)|^2 \\
        &-\frac{m^4}{2(2\pi)^2}\int d^3\hat{x}\sqrt{h}G_{Bb}(\hat{x};\vec{n}_1)G_{Bb}(\hat{x};\vec{n}_2)\int\widetilde{d^3k} \\
        &\times \,\mathrm{Im}\bigg[\int \widetilde{d^3p_1}\delta_+((m\hat{x}+k-p_1)^2+m^2)\\
        &\times\psi^*(m\hat{x},k)\psi(p_1,m\hat{x}+k-p_1) M_{p_1,m\hat{x}+k-p_1\to m\hat{x}k}\bigg] \\
        &+ \frac{m^4\pi^2}{2}\int d^3\hat{x}\sqrt{h}G_{Bb}(\hat{x};\vec{n}_1)G_{Bb}(\hat{x};\vec{n}_2) \\
        &\times \int \prod_{j=1}^4 \widetilde{d^3p_j}\psi^*(p_3,p_4)\psi(p_1,p_2)\delta_+((p_1+p_2-m\hat{x})^2+m^2) \\
        &\times M^\dagger_{p_3p_4\to m\hat{x},p_1+p_2-m\hat{x}}M_{p_1p_2\to m\hat{x},p_1+p_2-m\hat{x}}\delta^{(4)}(p_3+p_4-p_1-p_2) \\
        F_2^2 &= \frac{m^6}{2^{10}\pi^6}\int d^3\hat{x}_1\sqrt{h_1}d^3\hat{x}_2\sqrt{h_2}G_{Bb}(\hat{x}_1;\vec{n}_1)G_{Bb}(\hat{x}_2;\vec{n}_2) |\lbraket{\out}{\inte}{m\hat{x}_1,m\hat{x}_2}{\psi}|^2 \\
        &= \frac{m^6}{256\pi^6}\int d^3\hat{x}_1\sqrt{h_1}d^3\hat{x}_2\sqrt{h_2}G_{Bb}(\hat{x}_1;\vec{n}_1)G_{Bb}(\hat{x}_2;\vec{n}_2)|\psi(m\hat{x}_1,m\hat{x}_2)|^2 \\
        &-\frac{m^6}{128\pi^5}\int d^3\hat{x}_1\sqrt{h_1}d^3\hat{x}_2\sqrt{h_2}G_{Bb}(\hat{x}_1;\vec{n}_1)G_{Bb}(\hat{x}_2;\vec{n}_2) \\
        &\times \mathrm{Im}\left[\int \widetilde{d^3p_1}\delta_+((m\hat{x}_1+m\hat{x}_2-p_1)^2+m^2) \psi^*(m\hat{x}_1,m\hat{x}_2)\psi(p_1,m(\hat{x}_1+\hat{x}_2)-p_1) \right. \\
        &\left.\times M_{p_1,m(\hat{x}_1+\hat{x}_2)-p_1\to m\hat{x}_1m\hat{x}_2}\right] \\
        & +\frac{\pi^2m^6}{4}\int d^3\hat{x}_1\sqrt{h_1}d^3\hat{x}_2\sqrt{h_2}G_{Bb}(\hat{x}_1;\vec{n}_1)G_{Bb}(\hat{x}_2;\vec{n}_2) \\
        &\times \int\prod_{j=1}^4 \widetilde{d^3p_j}\psi^*(p_3,p_4)\psi(p_1,p_2)M^\dagger_{p_3p_4\to m\hat{x}_1m\hat{x}_2}M_{p_1p_2\to m\hat{x}_1m\hat{x}_2} \\
        &\times \delta^{(4)}(p_3+p_4-p_1-p_2)\delta^{(4)}(m\hat{x}_1+m\hat{x}_2-p_1-p_2).
\end{split}
\end{equation}
\bibliographystyle{JHEP}
\bibliography{references}


\end{document}